\documentclass[12pt,preprint]{aastex}
\begin{document}
\title{You Can't Get There From Here: Hubble Relaxation in the Local Volume}
\author{Alan B. Whiting}
\affil{Cerro Tololo Inter-American Observatory}
\affil{Casilla 603, La Serena, Chile}
\email{awhiting@noao.edu}

\begin{abstract}
A beginning end-point for galaxy motions within the 10-Mpc
Local Volume is constructed
by requiring a smooth distribution of (luminous) matter at
the time of recombination, which is shown to be equivalent to
a smooth Hubble flow at early times.  It is found, by this
purely kinematical method, that present
peculiar motions are too small by a factor of at least several 
(and largely in the wrong direction) to have produced the observed
structures within the age of the universe.
Known dynamical effects are inadequate to remove the discrepancy.
This result is different in origin from previously known ``cold
flow'' problems.
The simple dynamical picture often used within the Local Volume
(for instance, in deriving masses through calculation of a 
zero-velocity surface) is thus called into question.  
The most straightforward
explanation (though not the only possible) is that there exists
a large quantity of baryonic matter in this region so
far undetected, and unassociated with galaxies or groups.
\end{abstract}

\section{Introduction: Dealing with End-Points}

\begin{figure}
\plotone{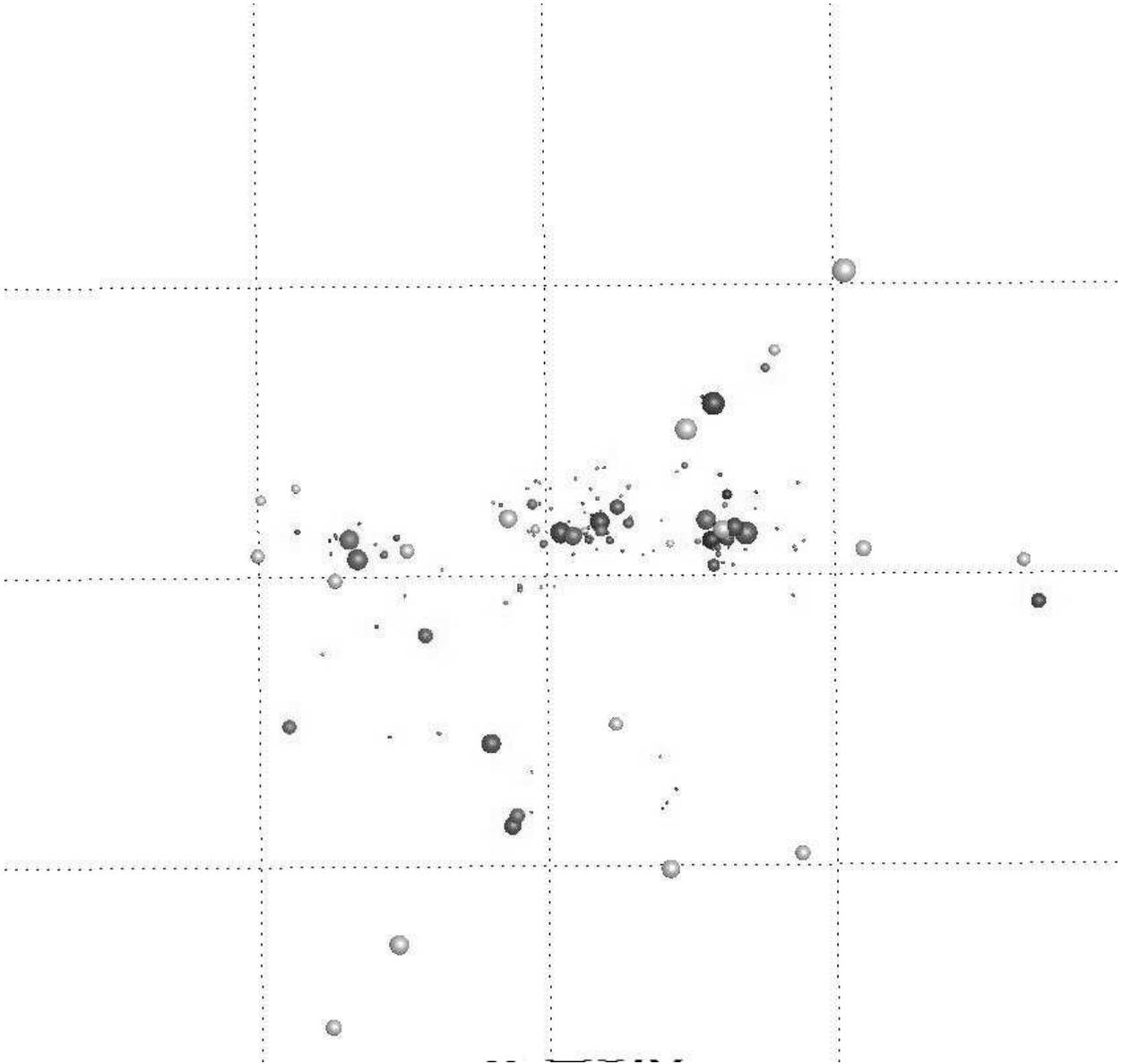}
\caption{A representation of the Local Volume, looking in the
Supergalactic Plane toward $L = 90\arcdeg$.  Galaxies are shown as spheres with volume
proportional to luminosity.  In the center is the Local
Group; to the left Centaurus, to the right IC 342 and M81/2.
The Supergalactic Plane itself is evident, as is the tendency
for galaxies to collect in groups.
Dotted lines are at 5 Mpc intervals.}
\label{volume}
\end{figure}

Matter was born smooth, and is everywhere in clumps.  Departures
from homogeneity in the CMB are tiny, on the order of a few dozen parts
per million; yet the Local Volume, the region within about ten
megaparsecs, is full of structure (see Figure (\ref{volume})).  Apart
from the galaxies themselves, there are groups of galaxies as
well as an overall flattening into the Supergalactic Plane.

The process by which a (nearly) smooth distribution of matter develops present-day
structure is complicated, engaging the efforts of many people (and computers).
But endpoints, if they can be usefully fixed, by themselves tell something about
the process on the average.  This study constructs a beginning endpoint for the
galaxies in the Local Volume and investigates the consequences of it.
(Structure formation by gravitational instability of small perturbations
actually starts at recombination, many years after the
Big Bang itself; but for our purposes it's close enough that it may be placed
at $t = 0$.)

\section{Constructing the Beginning}

There are two conditions we may impose on the Local Volume at the beginning.
\begin{itemize}
\item The motions of the galaxies must be {\em quiet}, fitting smoothly onto the
general Hubble flow at early times.  This condition is imposed, for instance, by \citet{P89}
in his Least Action calculations (and in subsequent studies using similar techniques).
It assumes, perhaps implicity, that each body retains some identity all the way back
to $t = 0$.
\item The material out of which the galaxies are made must be {\em smooth}, evenly
distributed (or with only very small fluctuations) over space.  This is a very direct interpretation
of measurements of the CMB: radiating matter at the surface of last scattering is
observed to have only small deviations from the average.
Note that this applies in the first instance to radiating baryonic 
matter alone; any effects of, or on, dark matter must be inferred.
\end{itemize}

These conditions are, strictly speaking, inconsistent.  The first requires galaxies,
or something like them, to exist at the beginning; the second requires that they do not.
But a reasonable compromise is possible.  We may assume, for the purposes of calculation,
that the material out of which a given galaxy is made comes from a simply-connected
region at the outset, which may be approximated by a body of the same total mass (or
total luminous mass) placed at its centroid.  The quality of the approximation will
depend on the shape of the region.  Strange shapes make it a poor assumption; but
very strange shapes seem unlikely.  In addition, in
their comparison of n-body simulations to the Least Action technique, \citet{DL95}
found that the shape of the original region had very little effect on subsequent
calculations.

The first condition is dynamic, and implementing it in any practical way would
involve us in difficult questions of mass-to-light ratio (and variations thereof
with time, mass, and possibly other parameters).  The second condition is purely
kinematic, much more robust and in fact what we will use for our analysis.  It
would be convenient to start with the``smooth'' condition and forget about 
quietness; unfortunately, the derivation of an important equation and the algorithm
we shall use depend on the dynamical formulation.  However, we shall show that,
with a trivial substitution, the end points are equivalent.  This is worth
emphasizing: dynamics is invoked and used at the outset, but the final result we
shall apply contains no dynamics at all.

\subsection{The Quiet Endpoint: Hubble Equilibrium}

Consider the motion of a body $i$ at a position ${\bf r}_i$ under
the action of $j$ other gravitating bodies of mass $m_j$, along with a possible
cosmological constant $\Lambda$.  Its acceleration is given by
\begin{equation}
\ddot{\bf r}_i = G \sum_{j \neq i} m_j \frac{{\bf r}_j - {\bf r}_i}
{|{\bf r}_j - {\bf r}_i|^3} + \frac{\Lambda}{3} {\bf r}_i
\end{equation}
in the nonrelativistic approximation (which holds for the situation of
interest to us here).
We convert to a different set of coordinates by factoring out a scalar
function $a$ which depends only on time, so that ${\bf r} = a {\bf x}$;
this gives us
\begin{equation}
\ddot{a} {\bf x}_i + 2 \dot{a} \dot{\bf x}_i + a \ddot{\bf x}_i =
\frac{1}{a^2} G \sum_{j \neq i} m_j \frac{{\bf x}_j - {\bf x}_i}
{|{\bf x}_j - {\bf x}_i|^3} + a \frac{\Lambda}{3} {\bf x}_i
\label{comoving1}
\end{equation}
where dots indicate derivatives with respect to time.
We are interested in comoving coordinates ${\bf x}$, that is, those
which are not functions of time.  In that case
\begin{equation}
{\bf x}_i = \frac{G}{a^2 \left( \ddot{a} - a \Lambda/3 \right)}
\sum_{j \neq i} \frac{{\bf x}_j - {\bf x}_i}{|{\bf x}_j - {\bf x}_i|^3}.
\label{heqm}
\end{equation}
This equation (which may be termed the equation of Hubble equilibrium), 
when applied at early times, is the mathematical formulation of the
quiet condition.  The configuration of bodies expands uniformly, at constant
comoving coordinates ${\bf x}_i$ in an undisturbed Hubble flow.  If 
such a configuration had survived to the present day, it would show
no peculiar velocities and no structure other than the bodies themselves.

A thorough study
of the solutions to Equation (\ref{heqm}) leads to some mathematically
interesting results, but most have no obvious application to cosmology.
A few are detailed in \citet{WH97}.
Note that it requires
that the expression involving $a$ in the denominator be a constant; this
results in the familiar expressions for the universal line-element in
(pressureless) Friedmann-Robertson-Walker universes.

To apply the quiet condition to a specific situation, we seek to produce
a Hubble equilibrium configuration from a given set of masses at given positions.
Now, one may think of the situation described by Equation (\ref{heqm}) as
an equilibrium between a set of inverse-square,
attractive gravitational forces due to the other
masses in the universe, and a general repulsive force linear in distance
form wherever we take as the center.  (Note that this general force is {\em not} due
to the cosmological constant; it is an artifact of conversion to a non-inertial
set of coordinates.)  As it stands the equilibrium is unstable: motion
toward a given mass makes the attraction to that mass stronger, resulting
in more motion toward that mass; moving away makes the gravitational
attraction weaker and the general repulsion stronger, increasing motion
away.  Using even a damped version of the dynamical equations, then,
will not find Hubble equilibrium configurations (nor even time-reversed
dynamical equations, since it can be shown that Equation (\ref{comoving1})
has growing modes in either direction).

However, a set of masses at arbitrary coordinates can be relaxed to
Hubble equilibrium using the ``treacle'' method suggested by \citet{LB95}.
From the equation of equilibrium (Eq. (\ref{heqm})) construct an effective
acceleration
\begin{equation}
{\bf A}_i = {\bf x}_i -G \sum_{j \neq i} 
\frac{{\bf x}_j - {\bf x}_i}{|{\bf x}_j - {\bf x}_i|^3}
\end{equation}
(where various negative constants have been absorbed into $-G$).  Since
the equilibrium is unstable this effective acceleration must be in a direction
away from equilibrium, so if we step
each body backwards along this acceleration a small amount,
we are getting closer to equilibrium.  We may then recalculate the accelerations and
iterate again.

It may be useful to think of Hubble relaxation sort of inside-out.  In
a line of thought attributed to Carlberg (but denied by him, at
least in this form),
consider that if all forces are
reversed, an equilibrium will remain an equilibrium but its stability
may change.  Thus if the cosmic repulsion of Equation (\ref{heqm}) is
replaced by a general drawing-in toward the center of the universe
and the gravitational attraction by a localized soft repulsion one
would still find the same equilibrium.  Think of drawing together a
number of soft foam balls in a fishing net\footnote{Something like this
was actually carried out, for other reasons and in another context,
by \citet{F70}.}.  This is what the the ``treacle'' relaxation
procedure does.

\subsection{The Smooth Endpoint}

We now compare the grainy universe of the last section with a completely
smooth one.
Consider a section of a universe of homogeneous, pressureless fluid of mass
density $\rho(t)$.  Assuming isotropy, we may determine its dynamics by looking only
at a sphere centered on our origin of coordinates.  We seek the equation of
motion of a point some vector distance ${\bf r}_i$ away from the origin.
This will be due to the combined gravitational attraction of all the
mass in our sphere, plus a cosmological constant if present:
\begin{equation}
\ddot{\bf r}_i = G \int \frac{\rho ({\bf r}') ({\bf r}'-{\bf r}_i)}
{\left| {\bf r}'-{\bf r}_i \right|^3} d{\bf r}' + \frac{\Lambda}{3} {\bf r}_i
\end{equation}
where the integral must formally be taken over all the sphere.  But note that
we need only consider the difference between a sphere of matter centered
on ${\bf r}_i$ (which would exert no net force) and this eccentric
sphere centered on the origin, which amounts to an extra shell on one side and a deficit on
the other.  If the reference sphere is taken to be large enough, these
shells can be treated as thin shells at a constant distance.

To remove any explicit dependence on the the details of our coordinate
system  we consider another
point a vector distance ${\bf r}_j$ from the origin, opposite ${\bf r}_i$,
and find
\begin{equation}
\ddot{\bf r}_i - \ddot{\bf r}_j =  G \int \frac{\rho ({\bf r}') ({\bf r}'-{\bf r}_i)}
{\left| {\bf r}'-{\bf r}_i \right|^3} d{\bf r}' -
 G \int \frac{\rho ({\bf r}') ({\bf r}'-{\bf r}_j)}
{\left| {\bf r}'-{\bf r}_j \right|^3} d{\bf r}'+ \frac{\Lambda}{3} \left( {\bf r}_i
- {\bf r}_j \right).
\end{equation}
Now we impose a comoving coordinate system, so that ${\bf r} = a {\bf x}$ and
the time derivatives of ${\bf x}$ vanish (that is, the smooth universe is in
uniform Hubble-flow expansion); this gives us
\begin{equation}
{\bf x}_i - {\bf x}_j = \frac{G}{\left( \ddot{a} - \Lambda a/3 \right) a^2}
\int \rho({\bf x}') \left( \frac{{\bf x}' - {\bf x}_i}{\left| {\bf x}' - {\bf x}_i
\right|^3} - \frac{{\bf x}' - {\bf x}_j}{\left| {\bf x}' - {\bf x}_j \right|^3}
\right) d{\bf x}'
\label{smooth}
\end{equation}
where the integral is taken over the appropriate thin shells.  Performing these
integrals gives the equation of motion for the scale factor:
\begin{equation}
a^2 \left( \ddot{a} - \frac{\Lambda a}{3} \right) = - \frac{4 \pi G}{3} a^3 \rho
\label{scale}
\end{equation}
which, if $\rho$ is taken as varying with time in the appropriate way, gives
the familiar Friedmann-Robertson-Walker results.

Now we consider, instead of the fluid universe, one in which the mass is concentrated at
points, but which is still in Hubble equilibrium; that is, Equation (\ref{heqm})
holds for each mass-point.  We assume that, on some scale, the discrete universe
approaches a constant-density continuous universe, so that we may assign the
same $\rho$ to a sufficiently large volume of each.  We also assume that
the discrete universe approaches isotropy on this same large scale.  By applying
Equation (\ref{heqm}) to the Hubble equilibrium distance between two points we
have
\begin{equation}
{\bf x}_i - {\bf x}_j = \frac{G}{\left( \ddot{a} - \Lambda a/3 \right) a^2}
\left( \sum_{k \neq i} m_k \frac{{\bf x}_k - {\bf x}_i}{\left|
{\bf x}_k - {\bf x}_i \right|^3} -
\sum_{k \neq j} m_k \frac{{\bf x}_k - {\bf x}_j}{\left|
{\bf x}_k - {\bf x}_i \right|^3} \right).
\label{lumpy}
\end{equation}
Comparing this with Equation (\ref{smooth}) we note that the condition of
smoothness amounts to each sum approximating the corresponding parts of the integral; and
that the fact that the integral need only be taken over thin shells at
large distances, means the approximation need only hold at large distances
also.

We now seek to compare the gravitational effect of the two Hubble-equilibrium point
masses $m_i$ and $m_j$ on each other 
with that of the corresponding region in the smooth universe.  To isolate the effects
we are interested in, let us remove the effects of the distant shells of matter from
each universe (smooth and grainy), say by momentarily imposing just the necessary
anisotropy.  Then, for the discrete case,
\begin{equation}
{\bf x}_i - {\bf x}_j  = \frac{G}{\left( \ddot{a} - \Lambda a/3 \right) a^2}
 (m_i + m_j) \frac{{\bf x}_j - {\bf x}_i}{\left|
{\bf x}_j - {\bf x}_i \right|^3}. 
\end{equation}

Now we impose the same behavior of $a$
as in the smooth universe, using Equation (\ref{scale}):
\begin{equation}
 {\bf x}_i - {\bf x}_j  = \frac{m_i + m_j}{-4 \pi a^3 \rho/3}
\frac{{\bf x}_i - {\bf x}_j}{\left| {\bf x}_i - {\bf x}_j \right|^3} 
\end{equation}
that is,
\begin{equation}
\rho = \frac{m_i + m_j}{4 \pi a^3 \left| {\bf x}_i - {\bf x}_j \right|^3 / 3}.
\end{equation}
So if the two masses are found, one on the center and one on the surface
of a sphere of radius
$\left| {\bf r}_i - {\bf r}_j \right|$, as they are in Hubble equilibrium,
we recover the behavior of a (dynamically) smooth universe 
(Equation \ref{scale}): the masses $m$ are spread
out in such a way as to give a constant density, when averaged over
any volume containing two or more bodies.  (This is much stronger
than the earlier assumption of approaching a constant density at large
scales.)

\subsection{Quiet is also Smooth}

Now if we interpret the quantities $m$ in the ``treacle'' algorithm not as
masses, but as luminosities, we find that it allows us to spread
out the luminous matter in the smoothest possible way in which the galaxies retain an
identity at the beginning.  As noted, we have used dynamics only to throw it
away at the end.

There are three things to emphasize at this point:
\begin{itemize}
\item The ``treacle'' result is robust, in that very few assumptions are made
leading to it.  It directly applies the requirement of a smooth distribution
of luminous matter.
\item The process is {\em not} dynamical.  How material got from smooth to clumpy is
not assumed, inferred or calculated.  In particular, the role of dark matter is
left quite open.
\item The algorithm produces the closest end point (averaged over all the galaxies)
to the present configuration.  There are very many ways of arranging proto-galactic
masses which satisfy Equation (\ref{heqm}); it is not contended that ``treacle''
produces the exact one which led to the structures we now see.  In some possible
arrangements, the distance
a given galaxy travels from the beginning to the end (in some comoving sense, which
will be dealt with in detail below) will be smaller; but another galaxy's distance
will be greater.  The version calculated here forms an averaged lower limit.  
(If the Local Volume in fact is dynamically simple, still undergoing infall into
the various galaxy groups and the Supergalactic Plane, the ``treacle'' end point
should be close to the actual end point.)
\end{itemize}

\section{Constructing the Final Endpoint}

In order to fix the final end point, as well as to start the ``treacle'' process, we
need accurate distances and photometry for all or most of the galaxies within the Volume.
We will also be using peculiar velocities, which must be determined from some
kinematic model.  In what follows we use the data and model of \citet{WH04}.  References
to sources and details of calculations will be found in that paper; here an outline is
given.

High-accuracy distances for 149 galaxies were taken from the literature.  These are
mostly Cepheid and tip of the Red Giant branch (TRGB) distances, with uncertainties of
$\sim 10\%$ or better.  Another 21 galaxies with distances of lower quality ($\sim 20\%$), but
among the brightest, were added.  The total of 170 by no means exhausts the total
population of the Local Volume, which is estimated to be around 500 \citep{KK04}.
However, it does include all known galaxies in the Volume brighter than $M_B \sim -18.5$,
and thus accounts for all but a few percent of the luminosity field.  Put another
way, all the unincluded dwarfs together would amount to perhaps 2 or 3 giant galaxies
of $M_B \sim -21$.

There are certainly galaxies hiding behind the plane of the Milky Way which have not
found their way into any catalog, and so are not accounted for.  One could also consider the
possibility of a dark galaxy of the type reported by \citet{MD05}, detected (in the
Virgo cluster) only in
HI radio emission, and so also missing from local galaxy catalogs.  But as they
note, anything like their discovery within about 6 Mpc would have been detected
by the HIPASS survey (which is not blocked by the Galactic plane).  It is
unlikely that any {\em large} galaxy in the Local Volume is missing from our data
set.  To change the results of this study materially, there would have to be many
such objects.  (Of course galaxies {\em outside} the Volume are not included.  Their possible
influence is considered below.)

Radial velocities for the galaxy sample were also taken from the literature.  Most
are HI measurements, with uncertainties of a few km s$^{-1}$; a few optical
radial velocities, with uncertainties typically of $\sim 10$ km s$^{-1}$, are
included.  (In general, radial velocities are the most accurately determined
of the input data.)

From these data an isotropic expansion model was fit by a direct least-squares procedure.  
The result
has the expansion rate of 65.5 km$^{-1}$ Mpc$^{-1}$ and a motion of the Sun 
(relative to the average of all galaxies)
of 340 km s$^{-1}$ in the direction of Galactic longitude
$98\arcdeg$, latitude $12\arcdeg$.  None of these figures is at great variance
with similar determinations.  The
rms peculiar radial velocity, the deviation from this model, is 79 km s$^{-1}$, of
which about 8-10 km s$^{-1}$ is attributable to distance uncertainties.

It is worth emphasizing that the peculiar velocities thus derived are entirely
kinematic.  They do not depend on any dynamical model
used to describe the overall behavior of the Volume.

Photometric measurements of each galaxy in $B$ and $K$ were sought in the literature.
Three galaxies in the sample did not have $B$ photometry; almost two-thirds were missing
$K$ measures.  In all cases, though, the missing data applied to dwarf galaxies, whose
total contribution to the total luminosity is small.  The apparent magnitudes were
converted to absolute using the known distances.

\section{Results and Analysis}

\begin{figure}
\plotone{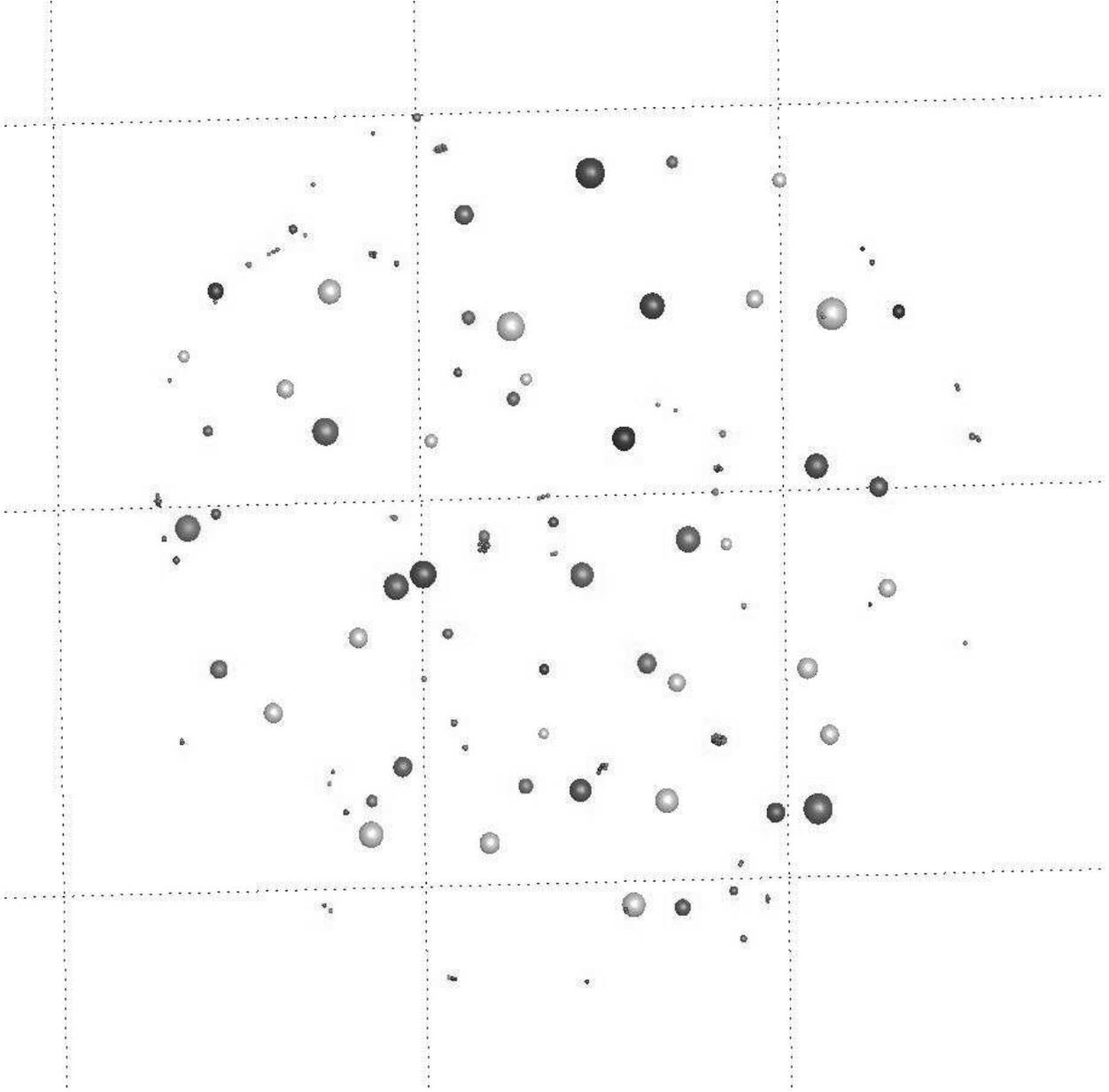}
\caption{The nearest Hubble equilibrium configuration as found by the 
``treacle'' method for the Local Volume.  As in Figure (\ref{volume}),
each galaxy is represented by a sphere of volume proportional to
luminosity.  The even distribution in three dimensions is only hinted at
by this two-dimensional view.}
\label{lvheqm}
\end{figure}

The treacle procedure was performed as outlined on the Local Volume data set,
and the resulting configuration scaled to be the same average size as the
input data.  (Since $K$-band luminosities are expected to be more representative of
luminous mass than $B$-band, the procedure was run using infrared photometry.)
A view of the resulting Hubble equilibrium configuration is shown in
Figure (\ref{lvheqm}).  The Supergalactic Plane and group structure
have been erased, as expected.  If we interpret the equation of Hubble equilibrium
dynamically (and if $K$ luminosity were exactly proportional to mass) 
this is one of the possible results if all the
Local Volume galaxies had expanded in an undisturbed Hubble Flow.  None of these
galaxies has any peculiar velocity, nor has any performed any peculiar motion.
If we interpret Hubble equilibrium as a kinematic condition on luminous baryons,
this configuration spreads them out in the smoothest possible way while each galaxy
retains its identity.

To derive any useful or quantitative results from this result, we need to
connect the endpoints.

We take, for convenience, the centroid of the Local Volume as the origin.
The average motion of a given
galaxy from $t=0$ to now is the vector from the origin to its present position,
${\bf r}_1$.  The average motion it would have followed in the absence of 
structure formation is the vector from the origin to its position in the Hubble
equilibrium configuration, ${\bf r}_2$.  The average motion required to get
from the smooth configuration to the real one is $\Delta {\bf r} = {\bf r}_2 - {\bf r}_1$;
we may call this the average peculiar motion.  (This is not expected to be
the actual peculiar motion, which will be curved in some way; but forms a
lower bound to it in distance.)  
To compare computed motions with anything observable we must take radial components,
so forming the computed average peculiar radial motion:
\begin{equation}
r_c = \Delta {\bf r} \cdot \hat{\bf r}
\end{equation}
which, when divided by the age of the universe, gives the computed average peculiar radial
velocity.  (The terms are clumsy, but it is necessary to be precise.)

\begin{figure}
\plotone{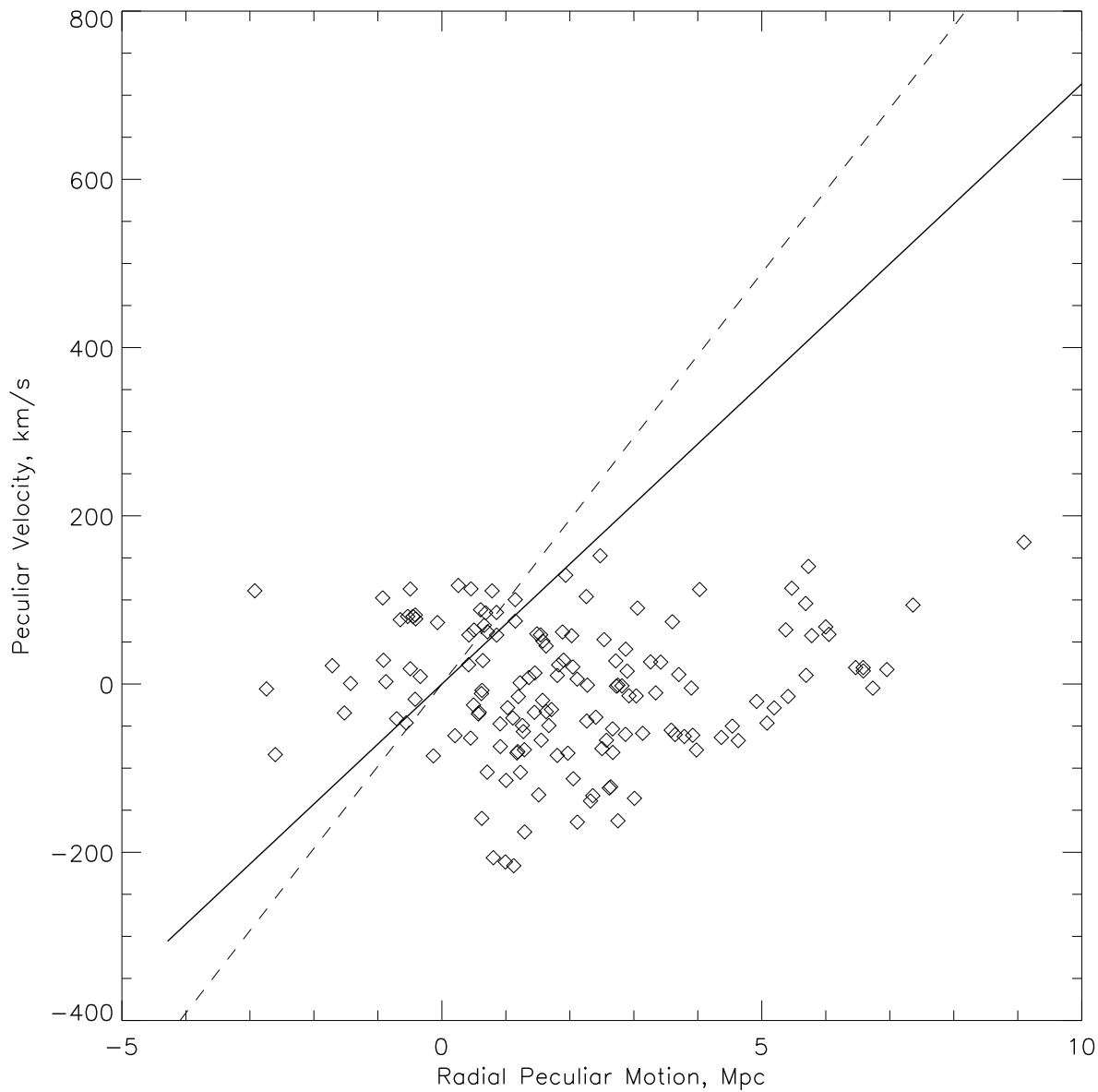}
\caption{Peculiar velocity for each body in the 149-galaxy sample of
\citet{WH04} plotted against the component of peculiar motion which
should show up as radial motion, assuming straight-line motion from
the nearest Hubble equilibrium configuration.  For an age of the
universe of 13.7 Gyr the solid line gives the most
direct relation between the two; for 10 Gyr, the dashed line.
Present peculiar velocities cannot account for more than about
2 Mpc of peculiar motion.}
\label{maine}
\end{figure}

A comparison between the calculated peculiar motions and observed peculiar
velocities
is shown in Figure (\ref{maine}).  (Only the 149 galaxies with high-accuracy
distances are used, since peculiar velocity uncertainties for the others were
much larger.)  The solid line gives the relationship one
would expect if all galaxies had proceeded directly from their places in the
beginning configuration to their present positions at a constant speed
over 13.7 Gyr.  That
is a simplistic idea and it is no surprise at all that the points do not lie
on that line.

But there is not even an average monotonic relationship here.
Galaxies are as likely to be heading in the wrong direction as
not, when compared to the assumption that they came from the nearest
smooth Hubble equilibrium configuration.  The assumption of
dynamical youth, that (apart, perhaps, from the very inner regions) all the
galaxy groups in the Local Volume are still experiencing infall and gathering
their outlying members from afar, doesn't appear to be true.  In particular, 
results from the timing
argument, whether in its original form by \citet{KW59} or in the zero-velocity
surface formulation of \citet{LB81}, must be reexamined (in spite of its
continuing popularity, of which \citet{VDB00} and \citet{K05} are examples).
Looked at another way, current peculiar velocities are in the wrong direction
to have moved the galaxies of Figure (\ref{volume}) more or less directly
from a smooth beginning endpoint.


For further
analysis we need to examine the relationship between the computed and
observed average speeds.

\subsection{Problems with Peculiar Speeds}

Leaving aside the matter of roundabout trajectories for the moment, we note that
the magnitude of the observed peculiar radial velocities is lower than the
required average peculiar radial velocities by a factor of three or more.  Of
course galaxies did not start with a certain speed on a certain course and
proceed thus from the beginning.  How should the present observed velocity,
$v_n$, compare with the average, $\bar{v}$?  Consider the following effects:

{\em Nonlinear evolution}.  Peculiar velocities are not produced from nothing.
They come, according to current theory (and there appears to be no viable
alternative), from gravitational effects of mass concentrations.  We expect
these to grow in the course of time, so their effect will be greater at
later times.  In addition, as objects get closer to each other their gravitational
influence will grow (until they pass; see next paragraph).  From each of these
considerations we expect $v_n > \bar{v}$, the opposite of what we see.

{\em Dissipationless collapse}.  Suppose that, instead of seeing a system in
the midst of infall, the components have interacted to the point of virialization.
How will the velocities compare?

For a uniform sphere of mass $M$ and total energy $-E$, the radius $r$ during
pressureless collapse is given by the parametric equations
\begin{eqnarray*}
r &= &\frac{GM}{E} \sin^2 (\eta) \\
t &=& \frac{GM}{\sqrt{2} E^{3/2}} \left( \eta - \frac{1}{2} \sin (2 \eta) \right)
\end{eqnarray*}
When the radius is just half of maximum, the time elapsed is
\begin{equation}
t_{1/2} = \frac{GM}{\sqrt{2} E^{3/2}} \left( \frac{\pi}{4} + \frac{1}{2} \right)
\end{equation}
Though the surface of the sphere has changed its position by half the original
radius, most points have not gone so far; the average motion over the volume
is 3/8 of the original radius.  With this figure and the time above we have
the expected average velocity.

Now consider the same sphere, but allow its components to virialize\footnote{In
what follows we are applying the Layzer-Irvine equation \citep{L63}, in which
the velocities are peculiar velocities.  For a truly virialized system, of
course, there is no overall expansion and all velocities are in a sense 
peculiar.  The Local Volume is clearly not virialized, since it is still
expanding.  The calculation is presented anyway, to show the 
effect of dynamical age and set a limit.}.  Then its
original total energy at maximum expansion, $-E = 3GM^2/(5R)$, is distributed
among its components such that
\begin{equation}
\frac{1}{2} M v_v^2 = E
\end{equation}
and the virialized radius is half the original one.  Comparing the average
virialized velocity
with the average infall velocity above, we find
\begin{equation}
\frac{v_v}{v_i} =5.7
\end{equation}
So dissipationless collapse will also tend to increase observed velocities.

{\em Dissipational collapse}.  In order to form galaxies at all, baryonic matter
must interact and lose kinetic energy to fall to the center of
dark-matter potential wells, so dissipation must act at some point in
structure formation.  However, when dealing with the whole Local Volume the
density of luminous matter is far lower than in galaxies (or protogalaxies).
Indeed, since the local Hubble flow is so close to the global value (and the age
is the same), the local average density must be close to the global density,
which is far too low for gas pressure to have any effect.  For dissipation to
happen galaxies must come together.

There is dissipation within galaxy groups.  Indeed, merging
has certainly taken place in the major galaxies (as the remnants of merged satellites
show).  That does not help with motions on several-Mpc scales, however.
To modify peculiar velocities sufficiently we need interactions between
galaxy groups, and they are too far away from each other now to have
interacted and then reached their present positions at their present speeds.

Possibly one could arrange trajectories so that galaxies coming from very different
directions or places preferentially merged, or interacted in some way so as to
convert most of their kinetic energy into other forms.  This does not seem likely
(merging is most probable for objects with similar velocities, not those speeding
past each other) and does not appear to be a feature of simulations.

Under expected conditions, then, dissipational collapse should not be an
important effect for motions on several-Mpc scales.

{\em The effect of external bodies}.  The 10 Mpc limit is somewhat arbitrary,
set by current observational
capabilities (resolved-star distances are very difficult to get at larger distances), not by
dynamical or kinematic considerations.  There are certainly galaxies beyond the border
which have some dynamical effect.

One effect could be to flatten the configuration of Figure (\ref{lvheqm}), which would
bring galaxies closer to their present positions in the Supergalactic Plane.  This
solves only a minor part of the problem, however; there are still gaps in the Plane which cannot
be filled at present peculiar velocities.  One could imagine placing external galaxies
in such a way as to fill in all the holes.  But this is terribly contrived, and very 
unlikely; and would result in some strange peculiar velocity patterns outside the Volume.

External masses could affect peculiar velocities.  But even large nearby clusters
do not generate the tidal fields of high spatial frequency needed to interfere with
Mpc-scale motions.  Virgo, for instance, can attract everything in the Volume
toward it (monopole effect) and add an ellipsoidal (quadrupole) component to peculiar
velocities.
But it cannot make a significant change in
the infall pattern into, say, the M81/2 group, which is at best
2 Mpc across, from 17 Mpc away.  And there does not appear to be an obvious way for
an external mass to have a general cooling effect on all peculiar motions.

{\em The decay of peculiar velocity}.  The peculiar velocity of an object moving
freely in an expanding universe decays, slowing down relative to the local
coordinate system.  As shown, for example, in \citet{P93}, $v \sim 1/a$, where
$a$ is the scale factor of the background cosmology.  For the simplest background,
the critical, lambda-less universe,
\begin{equation}
\frac{\bar{v}}{v_n} = \frac{t_n^{2/3}}{t_n-t_0} \int_{t_0}^{t_n} \frac{dt}{t^{2/3}}
\end{equation}
where variables with subscript $n$ denote the values now and subscript 0 the time
at which the particle began to propagate.  If we take the extreme case of $t_0=0$,
we get
\begin{equation}
\frac{\bar{v}}{v_n} = 3
\end{equation}
that is, the average speed of a freely-moving particle which starts out at the beginning
is three times its present speed.  For a flat universe with cosmological constant 
$\Lambda$ the expression is more complicated,
\begin{equation}
\frac{\bar{v}}{v_n} = \frac{\sinh(3 \sqrt{\Lambda} t_n /2)^{2/3}}{(3 \sqrt{\Lambda} /2)
(t_n-t_0)} \int_{t_0}^{t_n} \frac{dt}{\sinh(3 \sqrt{\Lambda} t /2)^{2/3}}
\end{equation}
but inserting the values appropriate for a 13.7 Gyr universe with a 70 km s$^{-1}$
Mpc$^{-1}$ Hubble constant we come up with
\begin{equation}
\frac{\bar{v}}{v_n} = 3.6
\end{equation}
(not an enormous difference).

This time we have a present velocity smaller than the average, and 
a glance at Figure (\ref{maine}) shows it to be apparently the right
amount.  If the magnitude of the peculiar velocities is increased by
a factor of three or four then the 13.7 Gyr line is found among them, not rising far
above and falling far below.  But there remains a problem: half of the galaxies
are still moving in the wrong direction.  They will need speeds of several times
that shown, in order to go around and come back (speaking roughly).  Even though
the decay of peculiar velocity acts in the right direction, it's still not
enough to fix the problem.

This is worth emphasizing.  By taking a galaxy at $t=0$ (before any had formed) and
giving it an initial push (before there was anything to push it), we have given it
more than the longest possible time for its peculiar velocity to decay; this is a limit.
Even under these conditions, it is not enough.

{\em Bias in the data}.  The result relies upon galaxy distances reported in the
literature.  They were chosen for study for a variety of reasons and with no necessary
relation to other objects, the opposite of a carefully-constructed survey.
In such a situation one expects to encounter bias, at least in the sense that
the data used are not strictly representative of the underlying population.
In the Local Volume in particular one expects more attention to be paid to
brighter galaxies (more interesting because more complicated, having a wider variety of internal
features, and of course easier to study).  And indeed the galaxies lacking in
the sample are mostly small dwarfs.

\begin{figure}
\plottwo{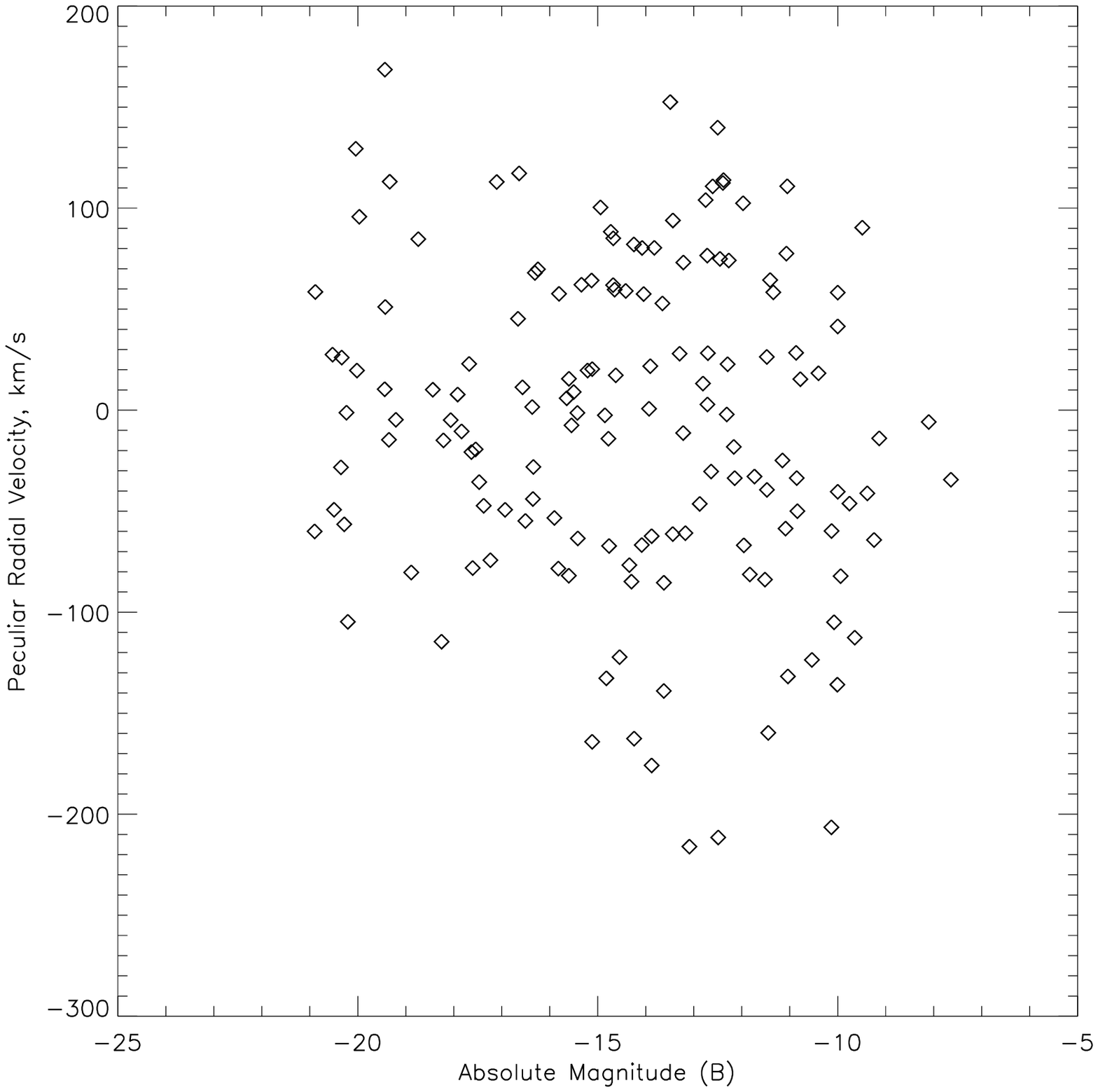}{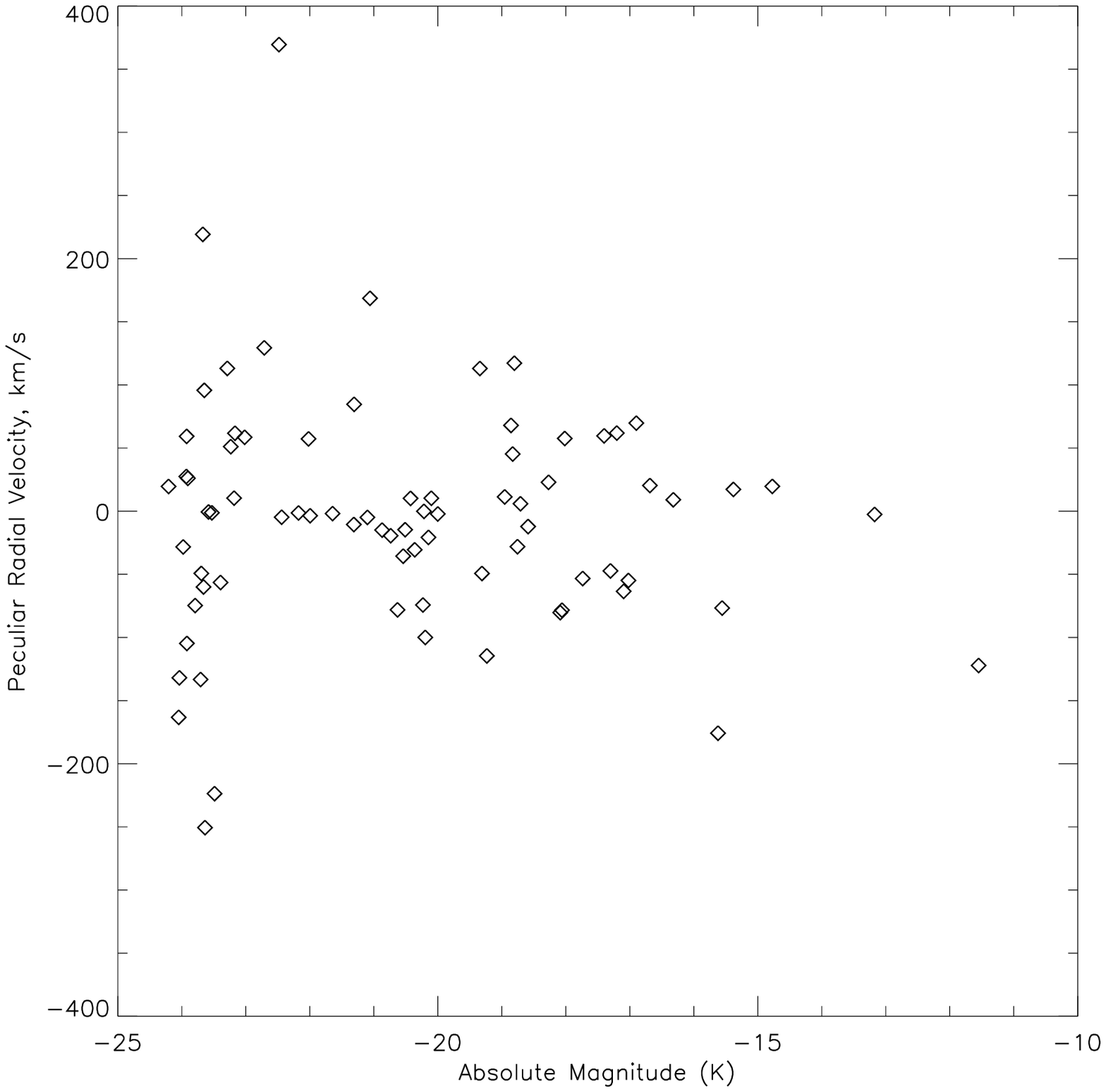}
\caption{Peculiar velocities for the sample of Local Volume galaxies, plotted
against (left) absolute magnitude in the $B$ band and (right) absolute magnitude
in the $K$ band.  Of the 149 galaxies with accurate distances, two did not have
$B$ photometry and do not show up on the left; about two-thirds lacked $K$
photometry and do not show up on the right.  The missing galaxies are dwarfs,
fainter than about -18 in $B$ and -15 in $K$.  Apart from an apparent lack
of high negative peculiar velocities among giant galaxies in $B$ and a
sparsely-populated plot for faint $K$, there are no visible correlations.}
\label{pvMK}
\end{figure}

But the sample does contain a wide range of luminosities, as shown in
Figure (\ref{pvMK}).  In addition, a bias in the sample toward bright galaxies seems to
make no difference in peculiar velocity; the graphs have a remarkably
even distribution.

And even if the Local Volume galaxies missing in the present sample had
radically different kinematics it would make no difference to the result,
since almost all of the luminosity is accounted for.  A few percent additional
would not affect the relaxed configuration in any significant way; it might
not even be measurable.

To dispose of another red herring,
dynamical bias has been reported in cosmological simulations between the
general dark matter velocity dispersion and those objects identifiable as
galaxies.  \citet{DM04} find that it amounts to 10-20\%, and so too small to
interest us here.  More importantly, as noted, Hubble relaxation using luminosity makes
no reference to dark matter at all.  Any bias between dark and luminous matter
is beside the point: the luminous matter by itself is required to be smooth
at $t=0$.

In summary, observed peculiar velocities are too small by a factor of at least a few
to produce the observed structure in the Local Volume from a smooth
distribution at recombination.  If we ignore all effects which tend to increase
current peculiar velocities relative to the overall time average, and push the
one decreasing effect to an unreasonable limit, 
we still cannot resolve the discrepancy.

Looked at another way, if the structures of Figure (\ref{volume}) are pushed apart at
the observed peculiar speeds over 13.7 Gyr, they would still be far from smoothly
distributed over the Volume.

\section{Comparison with Other Work}

\subsection{Cold Flow}

A point should be emphasized here: the low peculiar
velocities referred to here do {\em not} consitute the long-standing problem of
``cold flow.''  As found in the literature, that phrase seems to refer to three
slightly different situations:
\begin{itemize}
\item  Local peculiar velocities of
the same size as the observational uncertainties, about 60 km s$^{-1}$,
so that one could not
conclude that there was {\em any} departure from uniform expansion.
\citet{S86}, for example, makes this point.  This problem was in part due to small-number
statistics giving a slightly low dispersion and in part due to large
observational uncertainties.  As noted above, with uncertainties on
the 10 km s$^{-1}$ level, one can say that 70 km s$^{-1}$ of motion is real.
\item The discrepancy between simulations (mostly CDM with a critical,
lambdaless universe) giving peculiar velocity dispersions near 500
km s$^{-1}$ and local determinations one-fifth that size.  Going to a
different background cosmology seems to have resolved this issue;
see, for example, \citet{KH03}.  But, as noted, the present study
makes no dynamical calculations at all, so does not apply to this
problem.
\item The local velocity dispersion is still small compared with the
bulk motion of the whole Local Volume with respect to the CMB,
$\sim 600$ km s$^{-1}$.  Since I have not dealt with motion outside
the Volume, there is no direct connection between this ``cosmic Mach
number'' discrepancy and that shown by the Hubble relaxation process.
\end{itemize}

\subsection{N-Body Simulations}

Any new result in the area of structure formation prompts a call for comparison
with n-body simulations.  This is reasonable, given the great success the method
has enjoyed in producing results which match observations, especially on large
scales.  Unfortunately, there is no {\em direct} way to compare my procedure
with these codes.  Some of the assumptions (an expanding background universe,
a smooth beginning) are the same; but simulations are dynamical calculations,
dealing directly with the process of producing velocities, while Hubble
equilibrium (interpreted in the form of baryon luminosity) has nothing
to say about dynamics.  One cannot compare, for instance, methods
of dealing with dark matter, as one can between codes.

\subsubsection{Simulating the Local Volume?}

There are no doubt ways to make an indirect connection with some suitable
simulation of the Local Volume.  Unfortunately, such a phrase is something
of a misnomer: even if we had all the distance and velocity information to
start a backwards integration (which we certainly do not), the process is
mathematically unstable at early times, producing the wrong type of 
infinite motions.

N-body simulations work the other way, starting at some
early time and running forward.  The output is scrutinized for its
resemblence to the observed universe.  Overall features can be specified
to some degree \citep{KH03}, but in the strict sense no one simulates
the Local Volume; one simulates something which
shares the important features of the Volume (and what those features are will
vary from study to study).

Complicating the problem is the probability of the Volume being, in a dynamical
sense, rather unusual.  Over large volumes, for instance, the observed rms peculiar
velocity is much higher than locally, 400-500 km s$^{-1}$ \citep{HM03}, which one
might list as a fourth interpretation of ``cold flow''.  This means that a simulation
might be an accurate picture of the universe in general, containing all the right physics
and cosmological input parameters, while still failing to produce something {\em quite} like
the Local Volume.  Along these lines, note that
\citet{DM04} simulate the velocity dispersions well in
galaxy clusters but poorly (compared to observations) around individual galaxies.

There are also some difficulties in interpreting a simulation in terms of observational
data.  For instance, a recent simulation of the region around the Milky Way \citep{M05} 
has the right number of dark-matter haloes for large galaxies but far too many for
the smaller dwarfs (the venerable ``missing satellite'' problem).  To deal with this
kind of situation, the simulation of \citet{BF02} implements a relatively sophisticated
procedure for converting the characteristics of a dark halo into those of an observable 
galaxy, but one which they note has its limitations.

So producing a convincing simulation of the Local Volume is difficult.  Especially
problematic is judging the uncertainties in the procedure, so that one might be
able to say that something is indeed ruled out by the calculation.
The strategy of this paper is different.  Instead of searching for accurate models
it seeks to put limits on possible behavior.  The results are cruder, but more
robust.

\subsubsection{Testing Assumptions}

There has been one study testing some of the assumptions used
in this paper against simulations.  \citet{DL95} 
compared an n-body code with a Least Action reconstruction
technique similar to that in \citet{P89}.  It was a comparison between dynamical
calculations, with results framed in terms of the determined mean density
of the region ($\Omega$) in a lambdaless cosmology, and so dealing with
different questions than we are here.  However, the authors have two
results worth noting.  First they found that the
initial shape of a protogalaxy (the fact that it was not a point mass, as
assumed by the Least Action method) had little or no effect on any difference
in results.  The two calculations did indeed come up with different answers
for trajectories and velocities;
the authors attributed the difference to ``orphan'' dark matter particles not included
in identified galaxies.  As far as that is applicable, it underlines the
difficulty in deciding how to choose observables from a given simulation.

\subsection{The Baryon Deficit}

The main result of this study, that the luminous matter
in the Local Volume is moving too slowly to have been smoothly distributed
at recombination, has one strightforward explanation for which there is some
independent observational evidence.  If all the luminous matter in the universe
(or a representative part thereof) is counted up, that is everything in stars and
stellar remnants, it only makes up a small part of the baryon density required
by Big Bang nucleosynthesis.  \citet{PS92} noted the discrepancy and
the slightly updated figures of \citet{FGH98} make stars and remnants
17\% of the required baryons, the rest being ionized gas.  Suppose such gas is
present in the Local Volume in large quantities and distributed differently
from galaxies.  In particular, if it is found preferentially away from the
structures in Figure (\ref{volume}) then the baryons within the structures
need not travel so far (in comoving terms); thus it might allow all baryons at 
recombination to present a smooth
front to the world, while allowing galaxies to move to their present
positions at a more stately pace.

In principle the gas might be detectable in X-rays, in emission or absorption.  Some galaxy groups of quite
modest mass (less than 100 km s$^{-1}$ dispersion) have an observed X-ray flux \citep{HPM05},
though none are in the Local Volume, and for those the inferred gas mass fraction
is on the order of 10\%.  It is unclear how much gas would be necessary to explain
the peculiar velocity deficit (there is no obvious way to run the ``treacle'' 
procedure with an unknown additional field of matter), but certainly it would be several times that
of the (visible-light) luminous matter.  It might soon be possible to detect
much more tenuous material \citep{Y03}.

\section{Summary and Conclusions}

Imposing an end condition on the visible matter in the Local Volume has
revealed a problem with the currently-visible structures: they are moving
too slowly and often in the wrong direction for the observed peculiar
velocities to have produced them from a smooth distribution
at the time of recombination.
Setting limits (rather than performing detailed calculations), it has been
found that no known dynamical process explains the deficit.  

The most straightforward explanation is that much or most of the baryonic
matter in the Volume is in the form of ionized gas, or for some other
reason not now presently
detectable; and that it is distributed largely in the places where galaxies
are not.  In that case it might be observed soon, as X-ray technology improves.

This is not the only possible explanation, however.  Some previously unsuspected
way of modifying peculiar velocities might be at work, slowing everything down
greatly.  This solution, though, is only speculative.   

Complicating the interpretation of this result is the fact that
the Local Volume is both dynamically unusual and the only region which
may be examined in this detail.  It is possible that any volume in which
the peculiar velocity field can be determined to an accuracy of, say,
20 km s$^{-1}$ and sampled to scales of well under one Mpc, will show
this kind of anomaly.  It is also possible that something is happening
here that does not occur in most of the universe.

It is certain, however, that the
assumption of simple trajectories for galaxies within the Volume, freely
employed up to this time, must be reexamined.  It might indeed be a reasonable
approximation on some scales and in some cases, but subsequent use of it
must be justified.

\acknowledgements

The author gratefully acknowledges the suggestions of Dr. Donald Lynden-Bell
which led to Hubble relaxation and ``treacle'' in the first place.

\end{document}